\documentclass[prd,aps,twocolumn,showpacs,nofootinbib,color]{revtex4-1}

\usepackage{amsfonts,color,graphicx,epsfig,amsmath,multirow}
\usepackage[colorlinks=true,linkcolor=blue,citecolor=blue,urlcolor=blue]{hyperref}
\usepackage{mathtools}
\usepackage{slashed}
\usepackage{graphicx,cleveref}
\usepackage{bm}
\usepackage{slashed}

\begin{document}

\title{The $B \to \rho $ helicity form factors within the QCD light-cone sum rules}

\author{Wei Cheng$^1$}
\author{Xing-Gang Wu$^1$}
\author{Rui-Yu Zhou$^1$}
\author{Hai-Bing Fu$^2$}

\address{$^1$Department of Physics, Chongqing University, Chongqing 401331, P.R. China}
\address{$^2$ School of Science, Guizhou Minzu University, Guiyang 550025, P.R. China}

\date{\today}

\begin{abstract}

We study the $B \to \rho$ helicity form factors (HFFs) by applying the light-cone sum rules up to twist-4 accuracy. The HFF has some advantages in comparison to the conventionally calculated transition form factors, such as the HFF parameterization can be achieved via diagonalizable unitarity relations and etc. At the large recoil point, only the $\rho$-meson longitudinal component contributes to the HFFs, and we have $\mathcal{H}_{\rho,0}(0)=0.435^{+0.055}_{-0.045}$ and $\mathcal{H}_{\rho,\{1,2\}}(0)\equiv 0$. We extrapolate the HFFs to physically allowable $q^2$-region and apply them to the $B \to \rho$ semileptonic decay. We observe that the $\rho$-meson longitudinal component dominates its differential decay width in low $q^2$-region, and its transverse component dominates the high $q^2$-region. Two ratios $R_{\rm low}$ and $R_{\rm high}$ are used to characterize those properties, and our LCSR calculation gives, $R_{\rm low}=0.967^{+0.308}_{-0.285}$ and $R_{\rm high}=0.219^{+0.058}_{-0.070}$, which agree with the BaBar measurements within errors.

\end{abstract}

\pacs{13.25.Hw, 11.55.Hx, 12.38.Aw}

\maketitle

\section{introduction}

The $B$-meson decays are important for precision test of standard model (SM) and for seeking of new physics beyond the SM. Within the framework of SM, they can be used to fix the masses and couplings of the basic particles, research the CP-violation phenomena, determine more precise values for the Cabibbo-Kobayashi-Maskawa (CKM) matrix elements, and etc, cf. Refs.\cite{Kennedy:1990ib, Shifman:1978bx, Cabibbo:1963, Kobayashi:1973, Buras:1983ap, Chetyrkin:1996vx, Ball:1998kk, Deshpande:1997rr}.

For the $B$-meson decays, one has to deal with the one-particle, the two-particle, and the three or more particle matrix elements. Those hadronic matrix elements are key components for extracting useful information on the underlying flavor transitions and studying the decay constants, the transition form factors (TFFs), the mixings and decay amplitudes. The $\gamma$-structures of those non-perturbative hadronic matrix elements can be decomposed into Lorentz-invariant structures by using covariant decomposition, leading to basic TFFs for various decay channels.

\begin{table}[htb]
\begin{center}
\begin{tabular}{ccl}
\hline
~~Matrix element~~ & ~~TFFs~~ & ~~~~~~HFFs~~ \\
\hline
$\begin{array}{c}\langle V|\bar{q}\gamma^\mu b|B\rangle
\\ \langle V|\bar{q}\gamma^\mu\gamma^5 b|B\rangle\end{array}$ &
$\begin{array}{c}V\\ A_0, A_1, A_2\end{array}$ &
$\left\}\begin{array}{c} \mathcal{H}_{\mathcal{V},0},~\mathcal{H}_{\mathcal{V},t} \\
\mathcal{H}_{\mathcal{V},1},~\mathcal{H}_{\mathcal{V},2}\end{array}\right.$ \\[5mm]
$\begin{array}{c}\langle V|\bar{q}\sigma^{\mu\nu}q_\nu b|B\rangle \\
\langle V|\bar{q}\sigma^{\mu\nu}\gamma^5 q_\nu b|B\rangle\end{array}$ &
$\begin{array}{c}T_1\\T_2, T_3\end{array}$ &
$\left\}\begin{array}{c}\mathcal{H}_{\mathcal{T},0} \\
\mathcal{H}_{\mathcal{T},1},~\mathcal{H}_{\mathcal{T},2}\end{array}\right.$ \\ \hline
\end{tabular}
\caption{The relations among the $B\to$ vector meson transition form factors (TFFs), the helicity form factors (HFFs), and the hadronic matrix elements.}
\label{tab:formfact}
\end{center}
\end{table}

Specifically, for the $B \to$ light vector meson decays, we need to deal with seven TFFs for the hadronic matrix elements~\cite{Fu:2014uea, Bharucha:2010im}, which are shown in Table \ref{tab:formfact}. For convenience, we also present the relations among the $B\to$ vector meson helicity form factors (HFFs) and the hadronic matrix elements in Table~\ref{tab:formfact}.

The $B\to$ light vector meson decays have been analyzed by various experimental groups, such as the BaBar collaboration~\cite{Lees:2012tva, delAmoSanchez:2010af}, the Belle collaboration~\cite{Sibidanov:2013rkk}, the LHCb collaboration~\cite{Aaij:2013qta, Aaij:2012cq}, the ATLAS collaboration~\cite{ATLAS:2013ola}, the CLEO collaboration~\cite{Behrens:1999vv}. On the other hand, the TFFs/HFFs for the $B \to$ light vector meson decays have been calculated under various approaches, such as the light-cone sum rules (LCSR)~\cite{svz, Ball:1997rj, Huang:1998gp, Ball:2004rg, Huang:2008zg, AKhodjamirian:2010, Fu:2014pba, Fu:2014cna, Cheng:2017bzz, Ahmady:2013cga, Straub:2015ica}, the lattice QCD (LQCD)~\cite{Lattice96:1, Lattice96:2, Lattice98, DelDebbio:1997nu, Lattice04, Horgan:2013hoa, Horgan:2013pva, Agadjanov:2016fbd}, the perturbative QCD (pQCD)~\cite{Kurimoto:2001zj, Chen:2002bq, Kurimoto:2002sb, Keum:2004is, Fan:2013qz}, or some Phenomenological model~\cite{Cheng:2017smj, Cheng:2017sfk}. Those approaches are complementary to each other, which are applicable for different $q^2$-region. The pQCD approach is valid in low $q^2$-region, the LCSR is applicable in small and intermediate $q^2$-region around $m_b^2-2m_b\chi$ ($\chi\sim 500$ MeV is the typical hadronic scale of the decay) and the LQCD is applicable in high $q^2$-region. Among them, the LCSR prediction can be extrapolated to whole $q^2$-region, thus providing an important bridge for connecting various approaches.

There are large differences for the predicted and measured $B\to\rho$ decay widths at the large $q^2$-region, c.f. Refs.\cite{delAmoSanchez:2010af, Ahmady:2013cga, DelDebbio:1997nu}. In the paper, we shall adopt the LCSR approach to recalculate the $B\to\rho$ hadronic matrix elements. In different to previous LCSR treatment~\cite{Fu:2014cna, Fu:2014pba}, we shall express the hadronic matrix elements by using the HFF with the help of the covariant helicity projection approach~\cite{Korner:1989qb}. The HFFs are also Lorentz-invariant functions which can be formally expressed as the linear combination of the usually adopted TFFs.

\begin{table}[htb]
\begin{center}
\begin{tabular}{c c c c c}
\hline
~Transition~& ~$J^P$~ & ~Mass (GeV) ~ & ~~~~HFFs~~  \\
\hline
                 & $0^-$ & 5.28  & $\mathcal{H}_{\mathcal{V},t}$  \\
  $b\to d$  & $1^-$ & 5.33  & $\mathcal{H}_{\mathcal{V},1}$  \\
                & $1^+$ & 5.72  & $\mathcal{H}_{\mathcal{V},0}$,~$\mathcal{H}_{\mathcal{V},2}$\\
\hline
\end{tabular}
\caption{The masses of low-lying $B_d$ resonances~\cite{Bharucha:2010im} and their relations to the HFFs, which are obtained by relating the dominant poles in the LCSRs to those low-lying resonances.}
\label{tab:reson}
\end{center}
\end{table}

There are some advantages for the use of HFF~\cite{Bharucha:2010im}: I) Dispersive bounds on the HFF parameterization can be achieved via the diagonalizable unitarity relations; II) There are relations between the HFFs and the spin-parity quantum numbers, especially when taking the heavy-quark and/or large-energy limit. Thus, they can be conveniently adopted for considering the contributions from the excited states. The relations among the HFFs and the low-lying states can be obtained by relating the dominant poles in the LCSRs to those low-lying resonances. We present the masses of low-lying $B_d$ resonances with explicit quantum numbers $J^P$ in Table~\ref{tab:reson}, which shall be used in our numerical calculations; III) The LCSRs for the $B\to V$ HFFs can be conveniently used for studying the polarized decay widths.

The remaining parts of the paper are organized as follows. In Sec.II, we give the calculation technology for the $B \to \rho$ HFFs within the LCSR approach. In Sec.III, we present the numerical results. By extrapolating those HFFs to the whole $q^2$-region, we study the properties of the $B$-meson semileptonic decay $B\to\rho\ell\nu_\ell$. Sec.IV is reserved for a summary.

\section{Calculation technology for the $B\to\rho$ HFFs}

As for the $B\to\rho\ell\nu_\ell$ semileptonic decays, we need to deal with the hadronic matrix element:
\begin{equation}
\sum\limits_{\alpha=0,\pm,t}\langle \rho(k,\varepsilon_\alpha (k))|\bar q \, \gamma_\mu(1-\gamma^5) \, b |B(p)\rangle. \label{HME}
\end{equation}
where $k=(k^0,0,0,|\vec{k}|)$, $\varepsilon_\alpha(k)$ are $\rho$-meson longitudinal ($\alpha=0$) and transverse ($\pm$) polarization vectors. In the $B$-meson rest frame with the $z$ axis along the $\rho$-meson moving direction, and we have
\begin{eqnarray}
\varepsilon _0(k) &=&  \frac{1}{m_{\rho}} (|\vec{k}| ,0,0,k^0),  \\
\varepsilon _\pm(k) &=&  \mp \frac{1}{{\sqrt 2 }}  (0,1, \mp i,0),
\end{eqnarray}
where $|\vec{k}|=\sqrt{\lambda}/2m_B$, $k^0 = {(M_B^2 + m_\rho ^2-q^2)}/{2m_\rho}$ with $q=p-k$, $\lambda = (t_{-} - q^2)(t_{+} - q^2)$ with $t_\pm=(m_B\pm m_\rho)^2$. The polarization vectors satisfy $k\cdot\varepsilon_\alpha(k)=0$.

As proposed by Ref.\cite{Korner:1989qb}, one can adopt the covariant helicity projection approach to study those hadronic matrix element (\ref{HME}). The off-shell $W$-boson has similar polarization vectors as those of $\rho$-meson, e.g. the off-shell $W$-boson with momentum $q=(q^0,0,0,-|\vec q\,|)$ are
\begin{eqnarray}
\varepsilon _0(q) &=&  \frac{1}{{\sqrt {q^2 } }} \, (|\vec q| ,0,0, -q^0 ), \\
\varepsilon _\pm(q) &=&  \mp \frac{1}{{\sqrt 2 }} \, (0,1, \mp i,0), \\
\varepsilon _t(q) &=& \frac{1}{{\sqrt {q^2 } }} \, q,
\end{eqnarray}
where $|\vec q|=|\vec k|$, $q^0 = {(M_B^2 - m_\rho ^2 + q^2)}/{2m_\rho}$, and the extra vector $\varepsilon _t(q)$ is the time-like polarization vector. The linear combinations of the transverse helicity projection vector $\varepsilon _ {\pm} (q)$ give
\begin{eqnarray}
 \varepsilon _1 (q)  &=& \frac{{\varepsilon _ - (q) - \varepsilon _ + (q)}}{{\sqrt 2 }} = (0,1,0,0), \\
\varepsilon _2 (q)   &=& \frac{{\varepsilon _- (q) + \varepsilon _ +(q)}}{{\sqrt 2 }} = (0,0,i,0).
\end{eqnarray}

Using the off-shell $W$-boson polarization vectors, one can project out the relevant HFFs from the hadronic matrix elements~\cite{Bharucha:2010im}
\begin{eqnarray}
\mathcal{H}_{\rho,\sigma}(q^2 )  &=& \sqrt{\frac{q^2}{\lambda}} \, \sum\limits_{\alpha=0,\pm,t}
{\varepsilon_\sigma^{*\mu}(q)} \times \nonumber\\
&& \langle \rho(k,\varepsilon_\alpha(k)) |\bar q \, \gamma_\mu(1-\gamma^5) \, b |B(p)\rangle,
\label{HFF:Definition}
\end{eqnarray}
where $q=p-k$. In the following, we shall not consider the time-like HFF ($t$), which can be treated by using the same way and has no contribution to semileptonic decay width due to chiral suppression.

Following the standard LCSR procedures~\cite{Ball:1998kk, Ball:2004rg, Khodjamirian:2006st}, we can derive the LCSRs for the $B\to\rho$ HFFs. We first define a two-point correlation function as
\begin{eqnarray}
\Pi_{\sigma}(p,q) &=& -i \sqrt{\frac{q^2}{\lambda}} \, \sum\limits_{\alpha=0,\pm,t} {\varepsilon_\sigma^{*\mu}(q)}\int d^4 x e^{iq\cdot x}  \nonumber \\
&&\times \langle \rho(k,\varepsilon_\alpha(k))|T\{j_{V-A,\mu}(x),j_B^\dag (0)\}|0\rangle ,  \label{correlators}
\end{eqnarray}
where the currents $j_{V-A,\mu}(x) = \bar d (x){\gamma _\mu }(1 - {\gamma _5})b(x)$ and $j_B^\dag (0)=i m_b \bar b(0) \gamma_5 q(0)$ which has the same quantum state of the $B$-meson with $J^{P}=0^-$, and $\sigma=(0,1,2)$.

In the time-like $q^2$-region, one can insert a complete series of the intermediate hadronic states in the correlator (\ref{correlators}) and single out the pole term of the $B$-meson lowest pseudoscalar,
\begin{widetext}
\begin{eqnarray}
\Pi _{\sigma}^{\rm H }&=&\sqrt{\frac{q^2}{\lambda}} \, \sum\limits_{\alpha=0,\pm,t}
{\varepsilon_\sigma^{*\mu}(q)}\frac{\langle \rho(k,\varepsilon_\alpha(k))
|\bar q \, \gamma_\mu(1-\gamma^5) \, b |B\rangle \langle B|\bar b i \gamma_5 q|0\rangle }{m_b[m_B^2 - (p  +  q)^2]}
\nonumber \\
&& + \sqrt{\frac{q^2}{\lambda}} \, \sum\limits_{\alpha=0,\pm,t} \sum\limits_{\rm H}
{\varepsilon_\sigma^{*\mu}(q)} \frac{\langle \rho(k,\varepsilon_\alpha(k))
|\bar q \, \gamma_\mu(1-\gamma^5) \, b |B^H\rangle \langle B^H|\bar b i \gamma _5 q |0\rangle }{m_b[m_{B^H}^2 - (p  +  q)^2]},
\end{eqnarray}
\end{widetext}
where $\langle B|\bar b i \gamma_5 q|0\rangle={m_B^2 f_B}/{m_b}$ with $f_B$ being the $B$-meson decay constant. By replacing the contributions from the higher-level resonances and continuum states with the dispersion relations, the invariant amplitudes can be rewritten as
\begin{eqnarray}
\Pi _{\sigma}^{\rm H } &=& \frac{m_B^2 f_B}{m_b[m_B^2 - (p  +  q)^2]} \mathcal{H}_{\rho,\sigma}(q^2) \nonumber\\
&& +\int_{s_0}^\infty  \frac{\rho_\sigma^{\rm H}}{s - (p  +  q)^2}ds + \cdots,
\end{eqnarray}
where $s_0$ stands for the continuum threshold parameter and the ellipsis is the subtraction constant or the finite $q^2$-polynomial, which has no contribution to the final sum rules. The spectral densities $\rho^{\rm H}_{\sigma}(s)$ can be approximated by using the ansatz of the quark-hadron duality~\cite{Shifman:1978by}, i.e. $\rho^{\rm H}_{\sigma}(s)= \rho^{\rm QCD}_{\sigma}(s)\theta (s-s_0)$.

In the space-like $q^2$-region, i.e. $(p+q)^2 - m_b^2 \ll 0$ and $q^2 \ll m^2_b$ for the momentum transfer, which correspond to small light-cone distance $x^2 \approx 0$, the correlator (\ref{correlators}) can be calculated by using the operator product expansion (OPE). By using the $b$-quark propagator given by Ref.\cite{Huang:1998gp}, we obtain
\begin{eqnarray}
\Pi_{\sigma} ^{\rm OPE }(p,q) &=& -i \sqrt{\frac{q^2}{\lambda}} \, \sum\limits_{\alpha=0,\pm,t} {\varepsilon_\sigma^{*\mu}(q)} \int \frac{d^4 x d^4 k}{(2\pi )^4} \frac{e^{i(q-k)\cdot x}}{m_b^2 - k^2}\nonumber\\
 && \big\{k^\nu \langle \rho(k,\varepsilon_\alpha(k))|{\rm T} \{ \bar d(x)\gamma _\mu \gamma _\nu \gamma _5 q(0) \}|0\rangle \nonumber \\
&& + {k^\nu }\langle \rho(k,\varepsilon_\alpha(k))|{\rm T}\{\bar d(x)\gamma _\mu \gamma _\nu {q}(0)\}|0\rangle  \nonumber\\
&& + m_b \langle \rho(k,\varepsilon_\alpha(k))|{\rm T}\{ \bar d(x) \gamma_\mu \gamma_5 q(0)\}|0\rangle \nonumber\\
&& - m_b\langle \rho(k,\varepsilon_\alpha(k))|{\rm T}\{ \bar d(x) \gamma_\mu q(0) \}|0\rangle +\cdots \big\}.
\end{eqnarray}
The nonlocal matrix elements can be expressed in terms of the $\rho$-meson LCDAs of various twists~\cite{Ball:2004rg, Ball:2007zt}, which are put in the Appendix.

The LCSRs for the $B \to \rho$ HFFs are then ready to be derived by equating the correlator in the time-like and space-like regions due to analytic property of the correlator in different $q^2$-regions. After applying the Borel transformation, which removes the subtraction term in the dispersion relation and exponentially suppresses the contributions from unknown excited resonances, we get the required LCSRs for the HFFs:
\begin{widetext}
\begin{eqnarray}
\mathcal{H}_{\rho,0} &=& \frac{m_\rho m_b(m_B^2-m_\rho^2-q^2)}{2\sqrt{\lambda} m_\rho f_B m_B^2}\int_0^1 du e^{\left( {m_{B}^2 - s(u)} \right) / M^2}  \bigg\{ \frac{ m_\rho f_\rho^\bot \cal C}{2u^2 m_\rho ^2}\Theta(c(u,s_0))\phi_{2;\rho}^\bot(u) +\frac{m_\rho f_\rho^\bot}{2u} \Theta(c(u,s_0))\nonumber\\
&&\times \psi_{3;\rho}^\|(u)+ \frac{ m_b f_\rho^\parallel}{u} \Theta \left(c \left( u,s_0 \right) \right) \phi_{3;\rho}^ \bot \left( u \right) - m_\rho f_\rho^\bot\bigg[ \frac{m_b^2{\cal C}}{8u^4M^4} \widetilde{\widetilde\Theta}(c(u,s_0)) + \frac{{\cal C}-2m_b^2}{8u^3M^2} \widetilde\Theta(c(u,s_0))\nonumber\\
&&- \frac{1}{8u^2}\Theta(c(u,s_0))\bigg]\phi_{4;\rho}^\bot(u) - \frac{m_b m_\rho^2 f_\rho^\parallel }{u^2 M^2}\widetilde \Theta \left( c\left( u,s_0 \right) \right)C_\rho(u) - m_\rho f_\rho^\bot\bigg[\frac{\cal C}{u^3M^2}\widetilde\Theta(c(u,s_0))-\frac{1}{u^2} \nonumber\\
&&\times \Theta(c(u,s_0))\bigg]I_L(u) - m_\rho f_\rho^\bot \bigg[\frac{2m_b^2}{2u^2M^2}\widetilde \Theta(c(u,s_0))+\frac{1}{2u} \Theta(c(u,s_0)) \bigg] H_3(u)\bigg\}+\int_0^1 dv \int_0^1 du \int_0^1 d {\mathcal D}\nonumber\\
&&\times e^{(m_{B}^2 - s(u)) / M^2}\frac{\widetilde\Theta(c(u,s_0))}{u^2 M^2} \frac{m_b m_\rho^2(m_B^2-m_\rho^2-q^2)} {24 \sqrt{\lambda}m_\rho f_B m_B^2} \bigg\{f_\rho^\bot \bigg[\widetilde {\Psi} _{4;\rho}^\bot (\underline \alpha ) - 12\bigg(\Psi _{4;\rho}^\bot (\underline \alpha )-2v \Psi _{4;\rho}^\bot (\underline \alpha )\nonumber\\
&&+2\Phi _{4;\rho}^{\bot(1)} (\underline \alpha ) -2\Phi _{4;\rho}^{\bot(2)} (\underline \alpha )+4v \Phi _{4;\rho}^{\bot(2)} (\underline \alpha )\bigg)\bigg]\bigg(m_B^2-m_\rho^2+2 u m_\rho^2\bigg)+ 2 m_b m_{K^*} f_\rho^\parallel \bigg(\widetilde {\Phi} _{3;\rho}^\parallel (\underline \alpha)\nonumber\\
&& + 12 \Phi _{3;\rho}^\parallel (\underline \alpha  )\bigg)\bigg\}-\frac{\sqrt{\lambda}m_\rho m_b}{4 m_\rho f_B m_B^2}\int_0^1 du e^{\left( {m_{B}^2 - s(u)} \right) / M^2} \bigg\{ \frac{m_\rho f_\rho^\bot}{u m_\rho ^2}\Theta(c(u,s_0))\phi_{2;\rho}^\bot(u)- \frac{m_b f_\rho^\bot}{u M^2} \widetilde\Theta (c(u,s_0))  \nonumber \\
&&\times\psi_{3;\rho}^{\|}(u) - \frac{m_\rho f_\rho^\bot}{4}\bigg[\frac{m_b^2}{u^3M^4} \widetilde{\widetilde\Theta}(c(u,s_0)) + \frac{1}{u^2M^2} \widetilde\Theta(c(u,s_0))\bigg]\phi_{4;\rho}^\bot(u)+ \frac{2 m_b f_\rho^\parallel }{u^2 M^2}\widetilde \Theta \left( c\left( {u,{s_0}} \right) \right)
\nonumber \\
&&\times A_\rho(u) - \frac{m_\rho^2 m_b^3 f_\rho^\parallel}{2u^4 M^6}\widetilde {\widetilde{\widetilde \Theta }}\left( c\left( u,s_0 \right) \right)B_\rho(u) + \frac{2 m_b m_\rho^2 f_\rho^\parallel }{u^2 M^4}\widetilde {\widetilde \Theta }\left( c\left( u,s_0 \right)\right)C_\rho(u) + 2 m_\rho f_\rho^\bot \nonumber\\
&&\times \bigg[\frac{{\cal C} - 2m_b^2}{u^3 M^4}\widetilde{\widetilde\Theta}(c(u,s_0)) - \frac{1}{u^2M^2}\widetilde\Theta (c(u,s_0))\bigg]I_L(u)- \frac{m_\rho f_\rho^\bot}{u M^2}\widetilde \Theta (c(u,s_0))H_3(u)\bigg\} - \int_0^1 dv \int_0^1 du\nonumber\\
&&\times \int_0^1 d {\mathcal D} e^{\left( {m_{B}^2 - s(u)} \right) / M^2} \frac{ \sqrt{\lambda} m_b m_\rho^2 f_\rho^\bot} {24m_\rho f_B m_B^2(m_B+m_\rho)} \frac{ m_B + m_\rho}{u^2 M^2}\widetilde\Theta(c(u,s_0))\bigg[\widetilde {\Psi} _{4;\rho}^\bot (\underline \alpha ) + 12\bigg( 2v \Psi _{4;\rho}^\bot (\underline \alpha )\nonumber\\
&&-\Psi _{4;\rho}^\bot (\underline \alpha )+(4v-2) \Phi _{4;\rho}^{\bot(1)} (\underline \alpha ) + 2\Phi _{4;\rho}^{\bot(2)} (\underline \alpha ) \bigg)\bigg],
\label{HFF BV0}
\end{eqnarray}
\end{widetext}
\begin{widetext}
\begin{eqnarray}
\mathcal{H}_{\rho,1} &=& \frac{\sqrt{2q^2} m_b }{2f_B m_B^2}\int_0^1 du e^{\left( {m_{B}^2 - s(u)} \right) / M^2} \bigg\{f_\rho^\bot \Theta(c(u,s_0))\phi_{2;\rho}^\bot(u)+\frac{m_\rho m_b f_\rho^\parallel}{2u^2 M^2}\widetilde \Theta \left( c\left( u,s_0 \right) \right) \psi _{3;\rho}^ \bot (u)  \nonumber\\
&& - \bigg[\frac{m_b^2}{u^2M^4}\widetilde {\widetilde\Theta}(c(u,s_0))+\frac{1}{uM^2}\widetilde\Theta(c(u,s_0))\bigg]
\frac{ m_\rho^2 f_\rho^\bot}{4}\phi_{4;\rho}^\bot (u)\bigg\}+ \int_0^1 dv \int_0^1 du \int_0^1 d {\mathcal D} e^{(m_{B}^2 - s(u)) / M^2} \nonumber\\
&& \times  \frac{ \sqrt{2q^2} m_\rho^2 f_\rho^\bot} {6(m_B+m_\rho)} \frac{\widetilde\Theta(c(u,s_0))}{u^2 M^2}\bigg[(2v-1)\widetilde {\Psi} _{4;\rho}^\bot (\underline \alpha ) + 12\bigg(\Psi _{4;\rho}^\bot (\underline \alpha )-2 (v-1)(\Phi _{4;\rho}^{\bot(1)} (\underline \alpha )-\Phi _{4;\rho}^{\bot(2)} (\underline \alpha ))\bigg)\bigg],
\label{HFF BV1}
\end{eqnarray}
\end{widetext}
\begin{widetext}
\begin{eqnarray}
\mathcal{H}_{\rho,2} &=& \frac{\sqrt{2q^2}m_\rho m_b}{\sqrt{\lambda}f_B m_B^2}\int_0^1 du e^{\left( {m_{B}^2 - s(u)} \right) / M^2}  \bigg\{ \frac{ m_\rho f_\rho^\bot \cal C}{2u^2 m_\rho ^2}\Theta(c(u,s_0))\phi_{2;\rho}^\bot(u) +\frac{m_\rho f_\rho^\bot}{2u} \Theta(c(u,s_0))\nonumber\\
&&\times \psi_{3;\rho}^\|(u)+ \frac{ m_b f_\rho^\parallel}{u} \Theta \left(c \left( u,s_0 \right) \right) \phi_{3;\rho}^ \bot \left( u \right) - m_\rho f_\rho^\bot\bigg[ \frac{m_b^2{\cal C}}{8u^4M^4} \widetilde{\widetilde\Theta}(c(u,s_0)) + \frac{{\cal C}-2m_b^2}{8u^3M^2} \widetilde\Theta(c(u,s_0))\nonumber\\
&&- \frac{1}{8u^2}\Theta(c(u,s_0))\bigg]\phi_{4;\rho}^\bot(u) - \frac{m_b m_\rho^2 f_\rho^\parallel }{u^2 M^2}\widetilde \Theta \left( c\left( u,s_0 \right) \right)C_\rho(u) - m_\rho f_\rho^\bot\bigg[\frac{\cal C}{u^3M^2}\widetilde\Theta(c(u,s_0))-\frac{1}{u^2} \nonumber\\
&&\times \Theta(c(u,s_0))\bigg]I_L(u) - m_\rho f_\rho^\bot \bigg[\frac{2m_b^2}{2u^2M^2}\widetilde \Theta(c(u,s_0))+\frac{1}{2u} \Theta(c(u,s_0)) \bigg] H_3(u)\bigg\}+\int_0^1 dv \int_0^1 du \int_0^1 d {\mathcal D}\nonumber\\
&&\times e^{\left( {m_{B}^2 - s(u)} \right) / M^2}\frac{\widetilde\Theta(c(u,s_0))}{u^2 M^2} \frac{\sqrt{2q^2}m_b m_\rho^2} {12\sqrt{\lambda} f_B m_B^2} \bigg\{f_\rho^\bot \bigg[\widetilde {\Psi} _{4;\rho}^\bot (\underline \alpha ) - 12\bigg(\Psi _{4;\rho}^\bot (\underline \alpha )-2v \Psi _{4;\rho}^\bot (\underline \alpha )\nonumber\\
&&+2\Phi _{4;\rho}^{\bot(1)} (\underline \alpha ) -2\Phi _{4;\rho}^{\bot(2)} (\underline \alpha )+4v \Phi _{4;\rho}^{\bot(2)} (\underline \alpha )\bigg)\bigg]\bigg(m_B^2-m_\rho^2+2 u m_\rho^2\bigg)+ 2 m_b m_\rho f_\rho^\parallel \bigg(\widetilde {\Phi} _{3;\rho}^\parallel (\underline \alpha)+ 12 \Phi _{3;\rho}^\parallel (\underline \alpha  )\bigg)\bigg\}, \label{HFF BV2}
\end{eqnarray}
\end{widetext}
where we have implicitly set the factorization scale as $\mu$. $\int d{\cal D}=\int d\alpha_1 d\alpha_2 d\alpha_3 \delta(1 - \sum \limits_{i \rm = 1}^{\rm{3}} {\alpha_i})$. ${\mathcal C}=m_b^2+u^2m_\rho^2-q^2$, ${\mathcal E} = m_b^2 - u^2 m_\rho^2 + q^2$, $\mathcal F = m_b^2 - u^2 m_\rho^2 - q^2$, ${\mathcal H} = q^2/(m_B^2 - m_\rho^2)$, $\mathcal Q = m_B^2 - m_\rho^2 - q^2$, $c(\varrho,s_0) = \varrho s_0 - m_b^2 + \bar \varrho q^2 - \varrho \bar \varrho m_\rho^2$ and $s(\varrho)=[ m_b^2 - \bar \varrho(q^2 - \varrho m_\rho^2)] / \varrho $ ($\varrho = u$) with $\bar \varrho = 1 - \varrho$. $\Theta(c(u,s_0))$ denotes the usual step function. $\widetilde \Theta(c(u,s_0))$ and $\widetilde {\widetilde \Theta}(c(u,s_0))$ can be obtained from the surface terms $\delta(c({u_0},{s_0}))$ and $\Delta (c({u_0},{s_0}))$, whose explicit forms have been given in Ref.\cite{Fu:2014uea}. The functions $A_\rho(u)$, $B_\rho(u)$, $C_\rho(u)$, $H_3(u)$ and $I_L(u)$ are defined as:
\begin{eqnarray}
A_\rho(u) =&&\int_0^u dv \left[ \phi _{2;\rho}^\| (v) - \phi _{3;\rho}^ \bot (v) \right],  \\
B_\rho(u) =&&\int_0^u dv \phi _{4;\rho}^\| (v),  \\
C_\rho(u) =&&\int_0^u dv \int_0^v {dw} \left[\psi _{4;\rho}^\|(w) + \phi _{2;\rho}^\|(w)\right. \nonumber\\
&&\left.- 2 \phi_{3;\rho}^\bot(w) \right], \\
H_3(u)=&& \int_0^u dv \left[\psi_{4;\rho}^{\bot}(v)-\phi_{2;\rho}^\bot(v)\right]
\end{eqnarray}
and
\begin{eqnarray}
I_L(u) =&&  \int_0^u dv \int_0^v dw \big[\phi_{3;\rho}^\|(w) -\frac{1}{2} \phi_{2;\rho}^\bot(w) \nonumber\\
&& -\frac{1}{2} \psi_{4;\rho}^\bot(w)\big].
\end{eqnarray}

\section{Numerical results}

\subsection{Input parameters and the HFFs}

We take the $\rho$-meson decay constants~\cite{Ball:2007zt}, $f_{\rho}^\bot=0.145(9)~{\rm GeV}$ and $f_{\rho}^\|=0.216(9)~{\rm GeV}$, the $b$-quark pole mass $m_b=4.80\pm0.05~{\rm GeV}$, the $\rho$-meson mass ${m_{\rho}} = 0.775~{\rm GeV}$, the $B$-meson mass $m_B = 5.279~{\rm GeV}$~\cite{Agashe:2014kda} and the $B$-meson decay constant $f_B=0.160\pm0.019{\rm GeV}$~\cite{Fu:2014pba}. The factorization scale $\mu$ is set as the typical momentum transfer of $B\to \rho$, i.e. $\mu\simeq (m_{B}^2-m_b^2)^{1/2} \sim 2~{\rm GeV}$, and we set its error as $\Delta\mu=\pm1~{\rm GeV}$~\cite{Ball:2004rg}.

\begin{table}[htb]
\centering
\caption{The $\rho$-meson LCDAs with different twist-structures, where $\delta \simeq m_{\rho}/m_b$ \cite{Ball:2004rg}.} \label{DA_delta}
\begin{tabular}{ cc c  c  }
\hline
  & ~~twist-2~~  & ~~twist-3~~ & ~~twist-4~~  \\
\hline
~~$\delta^0$~~   & $\phi_{2;\rho}^\bot$  &  / & $\Phi _{4;\rho}^{\bot(1)}$, $\Phi _{4;\rho}^{\bot(2)}$ \\
$\delta^1$         & $\phi_{2;\rho}^\|$ & $\phi_{3;\rho}^\bot$, $\psi_{3;\rho}^\bot$, $\Phi_{3;\rho}^\|$, $\widetilde\Phi_{3;\rho}^\|$  & / \\

$\delta^2$         & / & $\phi_{3;\rho}^\|$, $\psi_{3;\rho}^\|$ & $\phi_{4;\rho}^\bot$, $\psi_{4;\rho}^\bot$, $\Psi_{4;\rho}^\bot$, $\widetilde{\Psi} _{4;\rho}^\bot$ \\

$\delta^3$         &  / & / & $\phi_{4;\rho}^\|$, $\psi_{4;\rho}^\|$ \\
\hline
\end{tabular}
\end{table}

Up to twist-4 accuracy, the needed $\rho$-meson light-cone distribution amplitudes (LCDAs) are grouped in Table~\ref{DA_delta}, in which $\delta=m_{\rho}/m_b \sim 0.16$. Since the contributions from the twist-4 terms themselves are numerically small, we thus directly adopt the twist-4 LCDA model derived from the conformal expansion of the matrix element to do the numerical calculation~\cite{Ball:2007zt}. Contributions from the twist-3 LCDAs $\phi_{3;\rho}^\bot$, $\psi_{3;\rho}^\bot$, $\Phi_{3;\rho}^\|$ and $\widetilde\Phi_{3;\rho}^\|$ are suppressed by $\delta^1$ and the twist-3 contributions from the LCDAs $\phi_{3;\rho}^\|$ and $\psi_{3;\rho}^\|$ are suppressed by $\delta^2$. The 2-particle twist-3 LCDAs, i.e. $\phi_{3;\rho}^\bot$, $\psi_{3;\rho}^\bot$, $\phi_{3;\rho}^\|$ and $\psi_{3;\rho}^\|$, can be related to the twist-2 LCDAs $\phi_{2;\rho}^\|$ and $\phi_{2;\rho}^\bot$ via the Wandzura-Wilczek approximation~\cite{Wandzura:1977qf,Ball:1997rj}. The 3-particle twist-3 LCDAs are also numerically small and we shall adopt the models of Ref.\cite{Ball:2007zt} to do the calculation. The twist-2 LCDAs, $\phi_{2;\rho}^\|$ and $\phi_{2;\rho}^\bot$, can be derived by integrating out the transverse momentum dependence of the twist-2 light-cone wavefunction model constructed in Refs.\cite{BHL, Wu:2010zc, Fu:2014cna, Fu:2014pba, Wu:2013lga, Huang:1994dy, Cao:1997hw, Huang:2004su}. For convenience, we call it as the WH-DA model, which states
\begin{eqnarray}
&& \phi _{2;\rho }^\lambda (x,\mu_0)  =  \frac{{A_{2;\rho}^\lambda \sqrt {3x\bar x} {m_q}}}{{8{\pi ^{3/2}}\widetilde f_\rho ^\lambda b_{2;\rho}^\lambda }}[1 + {B_{2;\rho}^\lambda }C_2^{3/2}(\varsigma )] \nonumber\\
&& \times \left[ {{\rm{Erf}}\left( {b_{2;\rho}^\lambda \sqrt {\frac{{{\mu^2_0} + m_q^2}}{{x\bar x}}} } \right) - {\rm{Erf}}\left( {b_{2;\rho}^\lambda \sqrt {\frac{{m_q^2}}{{x\bar x}}} } \right)} \right], \label{DA:WH}
\end{eqnarray}
where $\lambda=\|$ or $\perp$, respectively. The reduced decay constants $\widetilde f_{\rho}^\perp=f_{\rho}^\perp/\sqrt{3}$ and $\widetilde f_{\rho}^\|=f_{\rho}^\|/\sqrt{5}$, $\varsigma=2x-1$, and the error function ${\rm Erf}(x) = \frac2{\sqrt \pi}\int_0^x e^{ - t^2}dt$. The lepton quark mass $m_q$ is usually taken as $0.3$ GeV and we vary it within the region of $[0.2, 0.4]$ GeV for its uncertainty. The parameters $A_{2;{\rho}}^\lambda$, $B_{2;{\rho}}^\lambda$ and $b_{2;{\rho}}^\lambda$ can be determined by using the usual constraints:
\begin{itemize}
\item The normalization condition, $\int \phi_{2;\rho}^{\lambda}(x)dx=1$;
\item The average of the squared transverse momentum, $\langle{\bf k}_\bot^2 \rangle_{\rho}^{1/2} = 0.37~{\rm GeV}$~\cite{Wu:2010zc, Huang:2013yya}.
\item The second Gegenbauer moments of the twist-2 LCDAs $\phi^{\perp}_{2;\rho}$ and $\phi^{\|}_{2;\rho}$, $a_2^\bot (1~{\rm GeV}) = 0.14(6)$ and $a_2^\| (1~{\rm GeV}) = 0.15(7)$~\cite{Ball:2007zt}.
\end{itemize}

\begin{figure}[htb]
\centering
\includegraphics[width=0.45\textwidth]{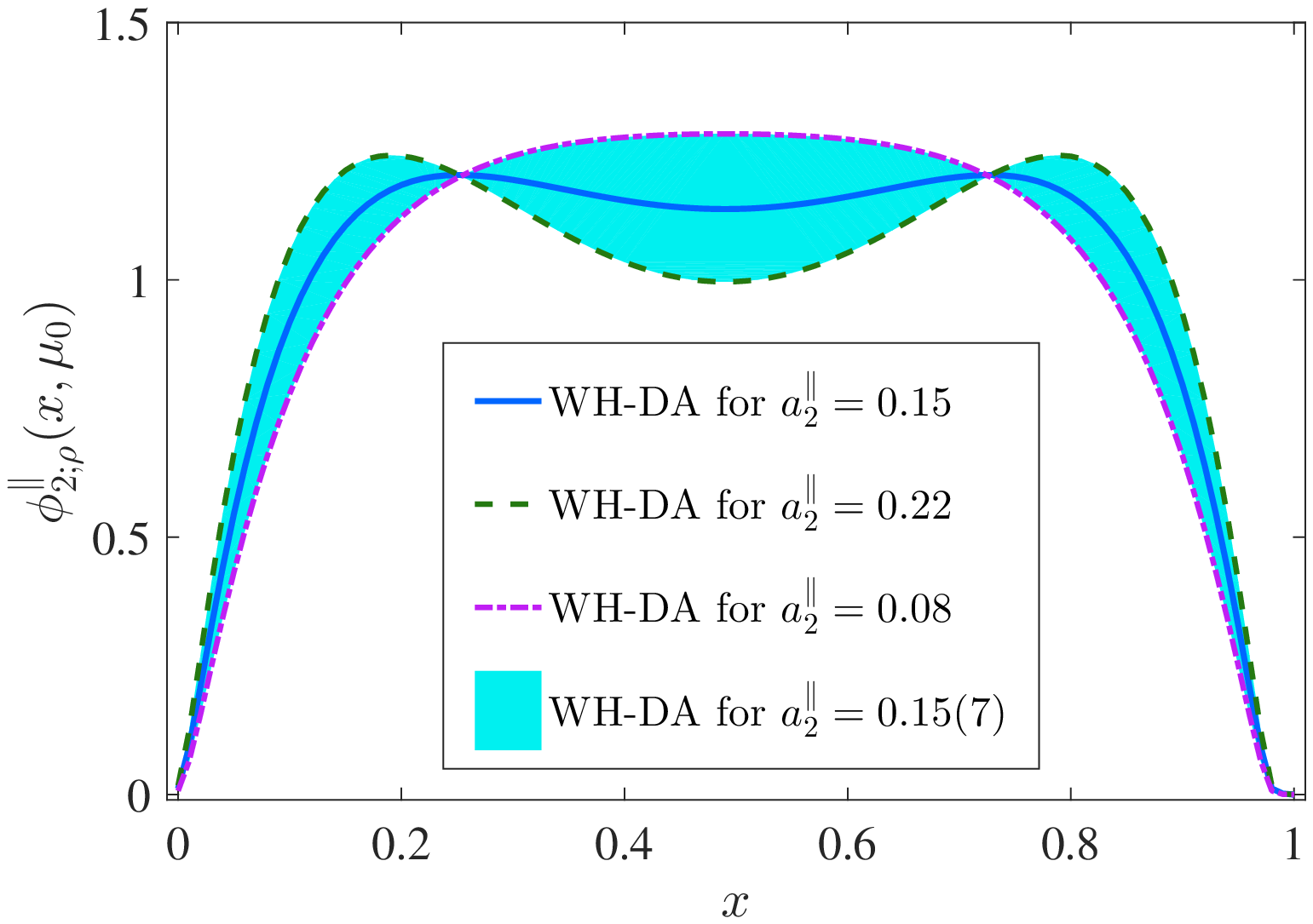}
\includegraphics[width=0.45\textwidth]{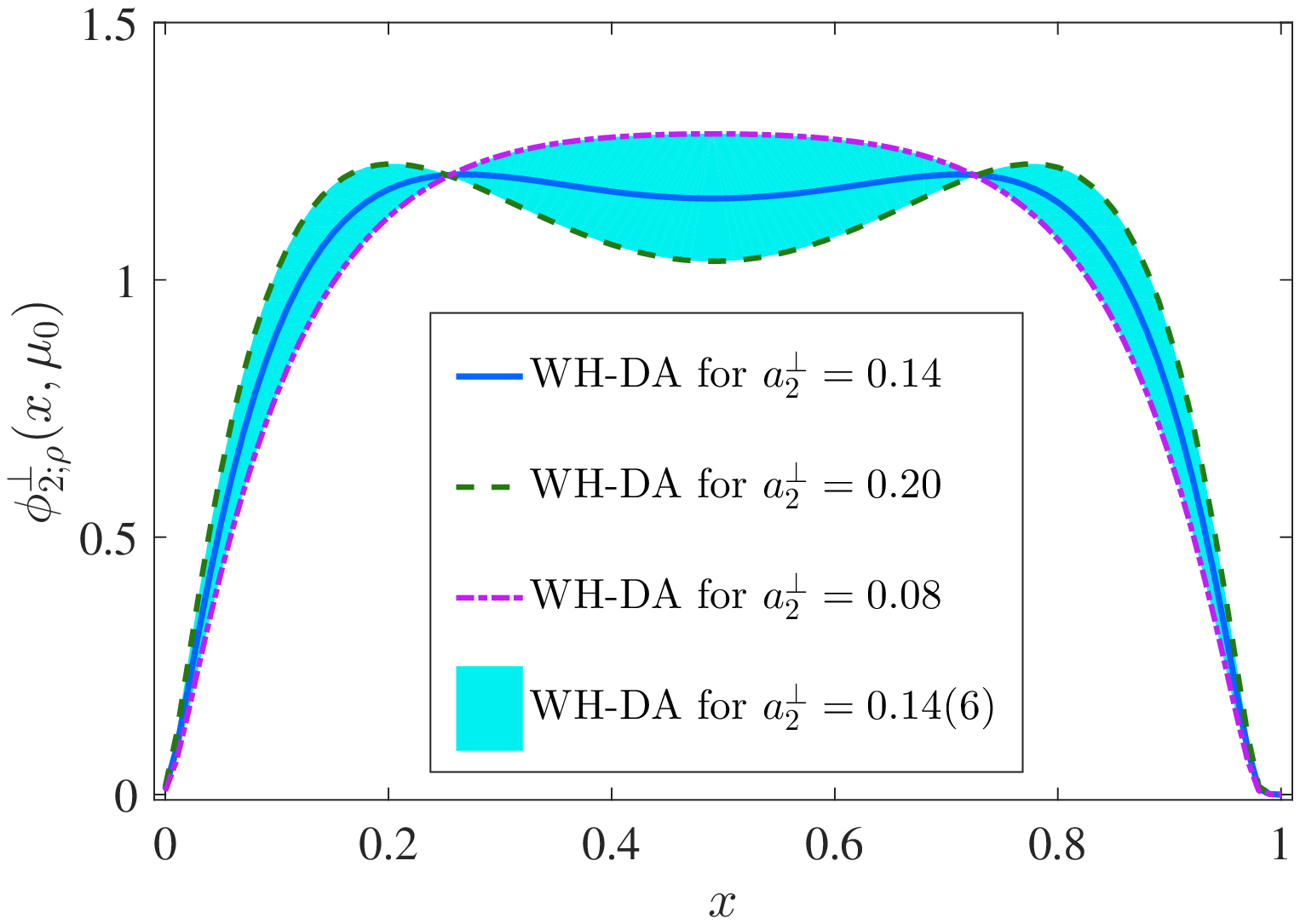}
\caption{The leading-twist LCDA $\phi_{2;{\rho}}^\lambda (x,\mu_0=1~{\rm GeV})$, where $\lambda$ stands for the transverse ($\lambda=\bot$) and longitudinal ($\lambda=\|$) components, respectively. $m_q=0.3$ GeV.}
\label{figDA}
\end{figure}

\begin{table}[htb]
\centering
\caption{Parameters of the ${\rho}$-meson transverse leading-twist LCDA for some typical choices of $a_{2}^{\bot}(1~{\rm GeV})$. $m_q=0.3$ GeV.}
\label{DA_parameter_1}
\begin{tabular}{c c c c}
\hline
~~$a_{2}^{\bot}$~~ & ~~$A_{2;\rho}^{\bot}$ ~~&~~ $B_{2;\rho}^{\bot}$ ~~&~~ $b_{2;\rho}^{\bot}$~~\\
\hline
$0.20$ & $22.679$& $0.151$ & $0.555$  \\
$0.14 $ & $23.808$ & $0.100$ & $0.572$ \\
$0.08$ & $25.213$& $0.050$ & $0.595$ \\
\hline
\end{tabular}
\end{table}

\begin{table}[htb]
\centering
\caption{Parameters of the ${\rho}$-meson longitudinal leading-twist LCDA for some typical choices of $a_{2}^{\|}(1~{\rm GeV})$. $m_q=0.3$ GeV.}
\label{DA_parameter_2}
\begin{tabular}{c c c c}
\hline
~~$a_{2}^{\|}$ ~~&~~ $A_{2;\rho}^{\|}$ ~~&~~$B_{2;\rho}^{\|}$ ~~&~~ $b_{2;\rho}^{\|}$~~\\
\hline
  $0.22$            & $22.620$          & $0.168$           & $0.549$ \\
  $0.15$            & $23.951$          & $0.109$           & $0.569$ \\
  $0.08$            & $25.275$          & $0.048$           & $0.590$ \\
\hline
\end{tabular}
\end{table}

Using those constraints, we can obtain the LCDA at the scale of $1$ GeV, whose behavior at any other scales can be achieved via the renormalization group evolution~\cite{Ball:2006nr}. The LCDA at any other scales can be obtained by using the conventional evolution equation. We present the parameters of $\phi_{2;\rho}^{\perp}$ and $\phi_{2;\rho}^{\|}$ in Table~\ref{DA_parameter_1} and \ref{DA_parameter_2}, and the corresponding curves in Fig.\ref{figDA}. Those two LCDAs are close in shape, both of which change from a convex behavior to a doubly humped behavior with the increment of the second Gegenbauer moment.

\begin{figure}[htb]
\centering
\includegraphics[width=0.45\textwidth]{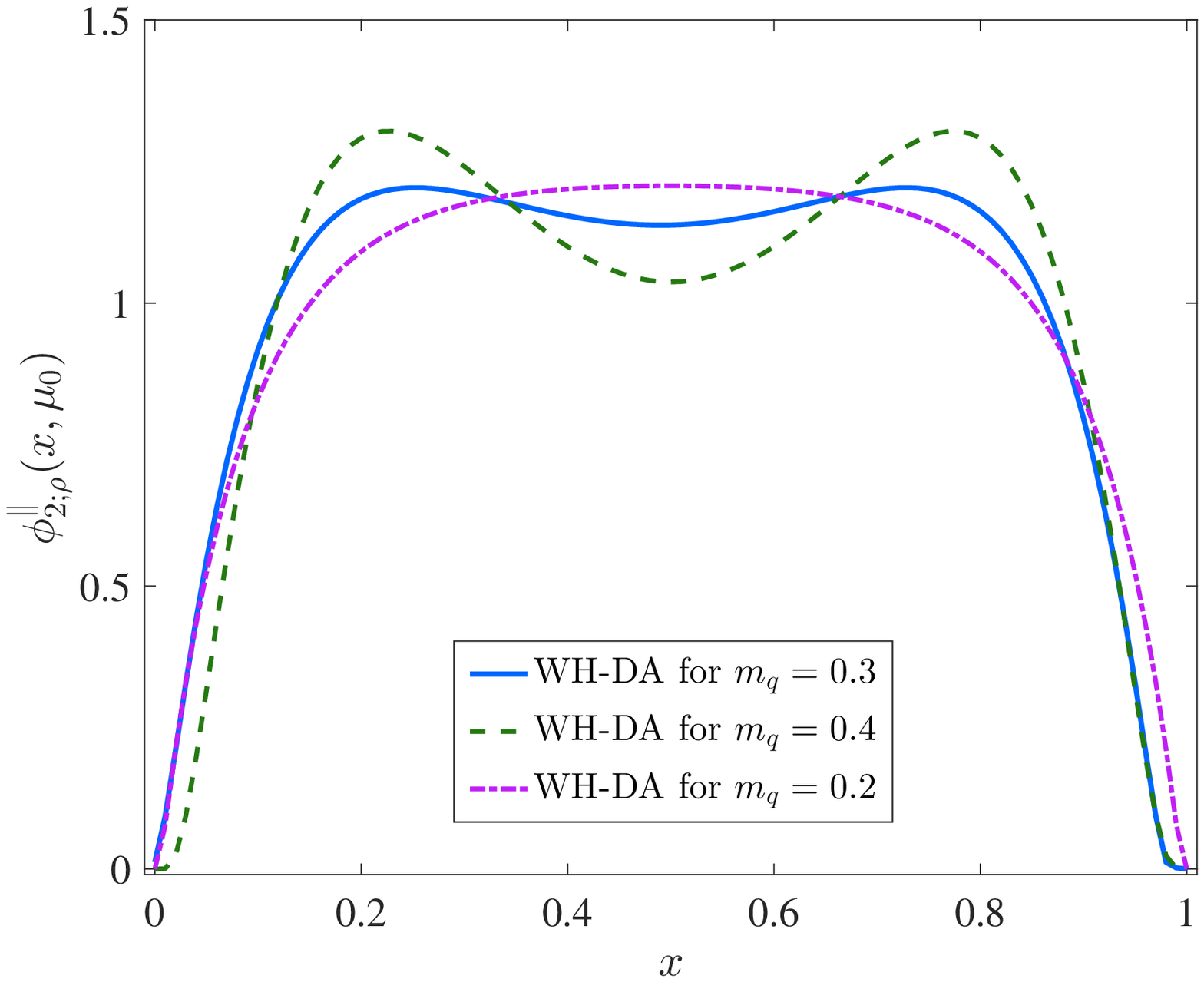}
\includegraphics[width=0.45\textwidth]{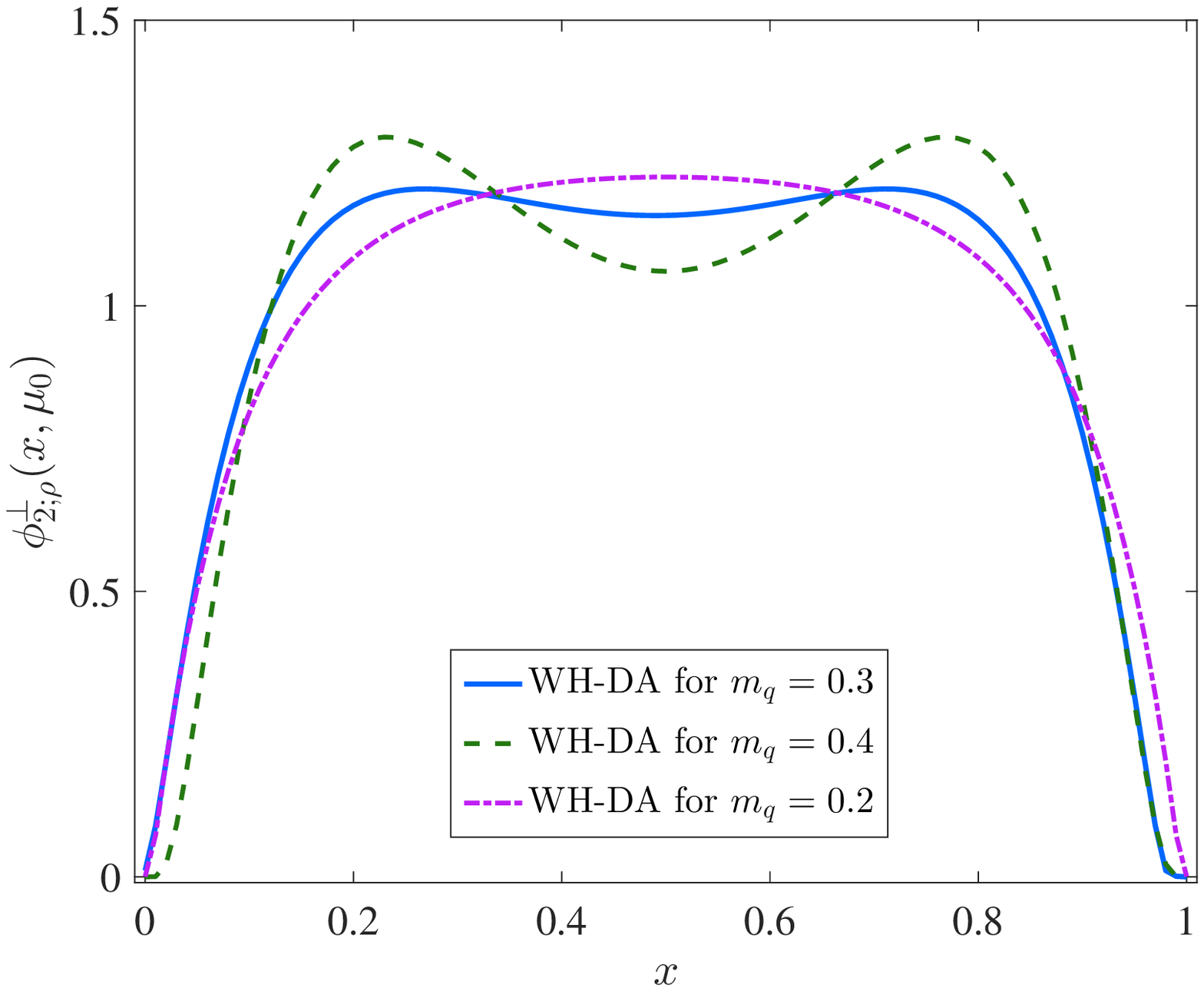}
\caption{The leading-twist LCDA $\phi_{2;{\rho}}^\lambda (x,\mu_0=1~{\rm GeV})$ for $m_q\in[0.2, 0.4]$ GeV, where $\lambda$ stands for the transverse ($\lambda=\bot$) and longitudinal ($\lambda=\|$) components, respectively. $a_2^\bot (1~{\rm GeV}) = 0.14$ and $a_2^\| (1~{\rm GeV}) = 0.15$. }
\label{figDAmass}
\end{figure}

Fig.\ref{figDAmass} shows how the LCDA  $\phi_{2;{\rho}}^\lambda$ changes with $m_q$. It is drawn by fixing all other input parameters to be their central values, and the LCDA parameters are refitted by fixing the second Gegenbauer moments $a_2^\bot (1~{\rm GeV}) = 0.14$ and $a_2^\| (1~{\rm GeV}) = 0.15$. As shown by Fig.(\ref{figDAmass}), different choices of light constitute quark $m_q$ can make sizable effects to the LCDA. Thus when discussing the uncertainties, the LCDA uncertainties from different choice of $m_q$ shall also be included.

\begin{table}[htb]
\caption{The Borel parameter $M^2$ for the HFFs $\mathcal{H}_{\rho,\sigma}$ at the continuum threshold $s_0=34.0$ $GeV^2$.}
\label{Borel}
\begin{tabular}{c c c c c c}
\hline
&$\mathcal{H}_{\rho,0}$ & $\mathcal{H}_{\rho,1}$ & $\mathcal{H}_{\rho,2}$
\\ \hline
$M^2$   & $25^{+0.5}_{-0.7}$ & $34.1^{+12.5}_{-7.8}$ & $21.8^{+3.7}_{-2.0}$
\\ \hline
\end{tabular}
\end{table}

As for the LCSRs of the HFFs, we also need to know the continuum threshold $s_0$ and the allowable range of the Borel parameter $M^2$, i.e. the so-called Borel window. The continuum threshold $s_0$, being as the demarcation of the $B$-meson ground state and higher mass contributions, is usually set as the one that is close to the first known resonance of the $B$-meson ground state. For the purpose, we set $s_0$ as $34.0\pm1.0$ ${\rm GeV}^2$, which indicates that the excitation energy is around 0.45 GeV to 0.65 GeV. The correlator is expanded over $1/M^2$, when we calculate it to all-power series, it shall be independent to the choice of $1/M^2$. However we only know its first several terms, and we have to set a proper range for $M^2$. As a conservative prediction, we require the continuum contribution to be less than $65\%$ of the total LCSR to set the upper limit of $M^2$, e.g.
\begin{eqnarray}
\frac{\int_{s_0}^\propto ds \rho^{\rm tot}(s)e^{-s/M^2}} {\int_{m_b^2}^\propto ds\rho^{\rm tot}(s)e^{-s/M^2}} \le 65\%. \label{con65}
\end{eqnarray}
Generally, the net contributions from the highest-twist terms increase with the decrement of $M^2$, and the lower limit of $M^2$ is usually fixed by requiring the highest-twist contributions to be small so as to ensure the convergence of the twist expansion. For the present considered three HFFs $\mathcal{H}_{\rho,\sigma}$, the twist-4 contributions behave quite differently. As a unified criteria for those HFFs, we adopt the flatness of the HFFs over $M^2$ to set the lower limit of $M^2$, e.g., we require the HFFs to be changed less than $1\%$ within the Borel window. The determined Borel window $M^2$ are listed in the Table~\ref{Borel}.

\begin{table}[htb]
\caption{Uncertainties of the LCSR predictions on the HFFs $\mathcal{H}_{\rho,\sigma}$ at the $q^2=10$ caused by the errors of the input parameters, e.g. $\Delta${DA} shows the uncertainty caused by varying the leading-twist LCDAs with the parameters listed in Tables \ref{DA_parameter_1} and \ref{DA_parameter_2}, in which the uncertainties caused by varying $m_q$ from $0.2 {\rm{GeV}} \to 0.4{\rm{GeV}}$ are also included. }
\label{HFF uncertainties}
\begin{tabular}{ c c c c c c c c }
\hline
&~~${\rm Central}$~~&~~$\Delta${DA}~~&~~$\Delta{\mu}$~~&~~$\Delta{M^2}$~~&~~$\Delta{s_0}$~~&~~$\Delta({m_b;f_B})$
\\ \hline
$\mathcal{H}_{\rho,0}$ & $0.688$ & $^{+0.003}_{-0.003}$ & $^{+0.000}_{-0.005}$ &$^{+0.006}_{-0.004}$& $^{+0.027}_{-0.027}$ & $^{+0.076}_{-0.062}$
\\
$\mathcal{H}_{\rho,1}$ & $0.314$ & $^{+0.002}_{-0.002}$ & $^{+0.000}_{-0.002}$ &$^{+0.000}_{-0.000}$& $^{+0.015}_{-0.018}$ & $^{+0.020}_{-0.016}$
\\
$\mathcal{H}_{\rho,2}$ & $0.408$ & $^{+0.003}_{-0.003}$ & $^{+0.000}_{-0.003}$ &$^{+0.003}_{-0.003}$& $^{+0.024}_{-0.026}$ & $^{+0.032}_{-0.025}$
\\
\hline
\end{tabular}
\end{table}

We take the HFFs $\mathcal{H}_{\rho,\sigma}(q^2=10)$ as explicit examples to show how the HFFs change with the input parameters. The results are collected in Table \ref{HFF uncertainties}, where errors from the $B$-meson decay constant $f_B$, the $b$-quark pole mass $m_b$, the $\rho$-meson mass ${m_{\rho}}$, the factorization scale $\mu$, the Borel parameter $M^2$ and the continuum threshold $s_0$. Table \ref{HFF uncertainties} shows that the main errors of those HFFs come from the parameters $m_b$, $f_B$, and $s_0$, whose effects could be up to $\sim 10\% -20\%$ accordingly.

\subsection{Extrapolation of the HFFs to all $q^2$-region}

The LCSR method is only valid for large energy of the final-state vector meson, e.g. $E_\rho \gg \Lambda_{\rm QCD}$. It implies a not too large $q^2$ via the relation $q^2 = m_B^2 - 2 m_B E_\rho$, e.g.
\begin{displaymath}
0\leq q^2\leq q^2_{\rm LCSR, max} \simeq 14~{\rm GeV}^2.
\end{displaymath}
On the other hand, the allowable physical range for $q^2$ is about $[0, 20.3]~{\rm GeV}^2$, in which the upper limit is fixed by $q^2_{\rm max}=(m_B-m_\rho)^2$~\cite{Ball:2004rg}. We adopt the method suggested by Ref.\cite{Bharucha:2010im} to do the extrapolation of the HFFs, i.e. the HFFs $\mathcal{H}_{\rho,\sigma}$ shall be extrapolated as a simplified series expansion as follows:
\begin{eqnarray}
\mathcal{H}_{\rho,0}(t) &=&\frac{1}{B(t) \sqrt{z(t,t_-)} \phi_T^{V-A}(t)} \sum_{k=0,1} a_k^{\rho,0} z^k ,  \\
\mathcal{H}_{\rho,1}(t) &=&\frac{\sqrt{-z(t,0)}}{B(t) \phi_T^{V-A}(t)} \sum_{k=0,1} a_k^{\rho,1} z^k ,   \\
\mathcal{H}_{\rho,2}(t) &=&\frac{\sqrt{-z(t,0)}}{B(t) \sqrt{z(t,t_-)} \phi_T^{V-A}(t)} \sum_{k=0,1} a_k^{\rho,2} z^k ,
\end{eqnarray}
where $\phi_I^X(t)=1$, $\sqrt{-z(t,0)}=\sqrt{q^2}/m_B$, $B(t)=1- q^2/m_{\rho,\sigma}^2$, $\sqrt{z(t,t_-)}=\sqrt{\lambda}/m_B^2$, and
\begin{eqnarray}
z(t)=\frac{\sqrt{t_+ - t}-\sqrt{t_+ - t_0}}{\sqrt{t_+ - t}+\sqrt{t_+ - t_0}}
\end{eqnarray}
with $t_\pm=(m_B\pm m_\rho)^2$ and $t_0=t_+(1-\sqrt{1-t_-/t_+})$.

\begin{table}[htb]
\centering
\caption{The fitted parameters $a^{\rho,\sigma}_{k}$ for the HFFs $\mathcal{H}_{\rho,\sigma}$, where all input parameters are set to be their central values.}
\label{analytic}
\begin{tabular}{ c  c c c c c c c }
\hline
&$\mathcal{H}_{\rho,0}$ & $\mathcal{H}_{\rho,1}$ & $\mathcal{H}_{\rho,2}$
\\ \hline
$a_0^{\rho,\sigma}$    & 0.257 & $0.386$ & $0.354$
\\
$a_1^{\rho,\sigma}$    & 1.511 & $-1.020$ & $-0.310$
\\
$\Delta$   & 0.238 & $0.045$ & $0.128$
\\ \hline
\end{tabular}
\end{table}

The parameters $a_k^{\rho,\sigma}$ can be determined by requiring the ``quality'' of fit ($\Delta$) to be less than one, where $\Delta$ is defined as
\begin{equation}
\Delta=\frac{\sum_t\left|\mathcal{H}_{\rho,\sigma}(t)-\mathcal{H}_{\rho,\sigma}^{\rm fit}(t)\right|} {\sum_t\left|\mathcal{H}_{\rho,\sigma}(t)\right|}\times 100, \label{delta}
\end{equation}
where $t\in[0,\frac{1}{2},\cdots,\frac{27}{2},14]~{\rm GeV}^2$. We put the determined parameters $a_{k}^{\rho,\sigma}$ in Table~\ref{analytic}, in which all the input parameters are set to be their central values.

\begin{figure}[htb]
\begin{center}
\includegraphics[width=0.45 \textwidth, height=0.25 \textwidth]{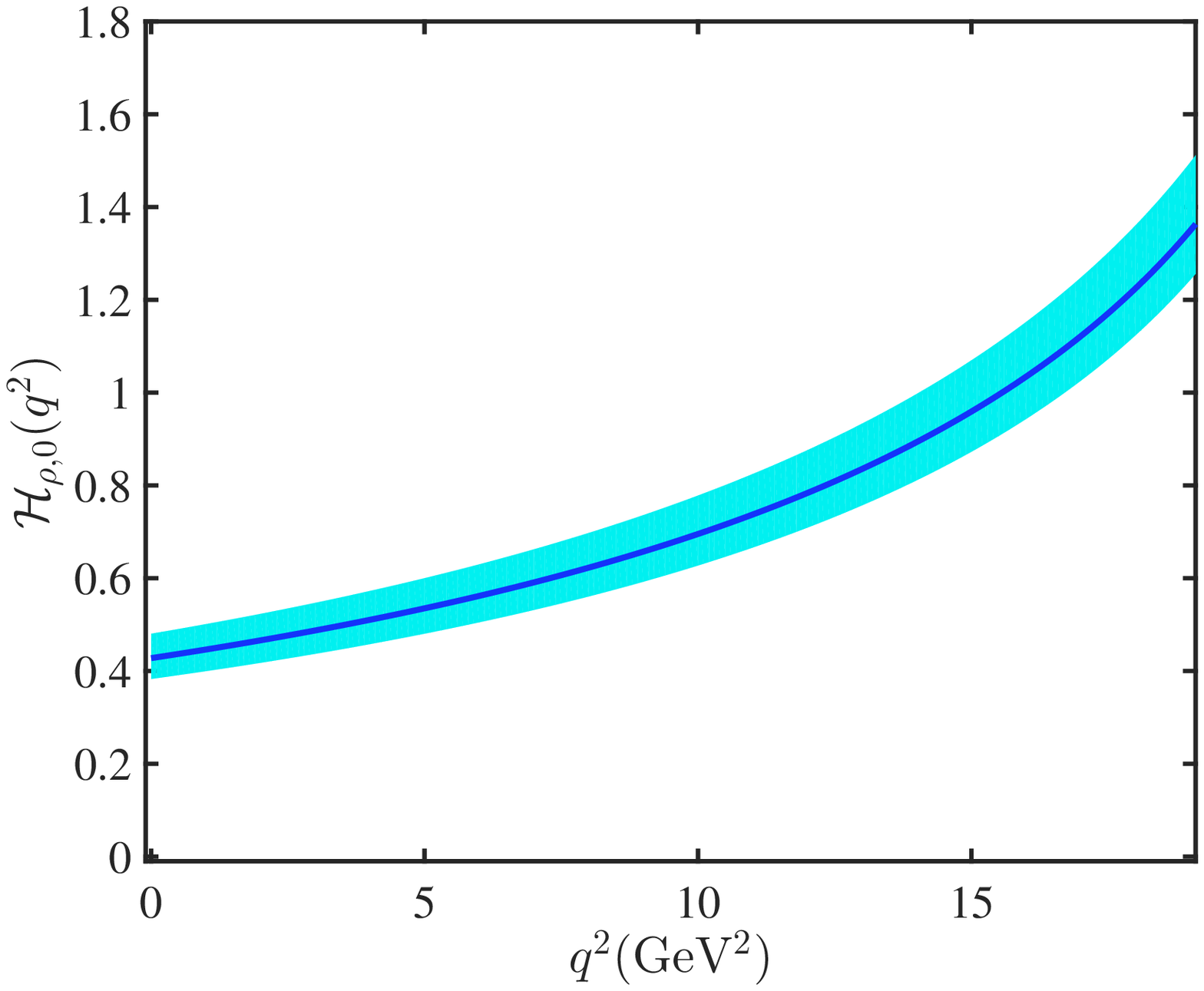}
\includegraphics[width=0.45 \textwidth, height=0.25 \textwidth]{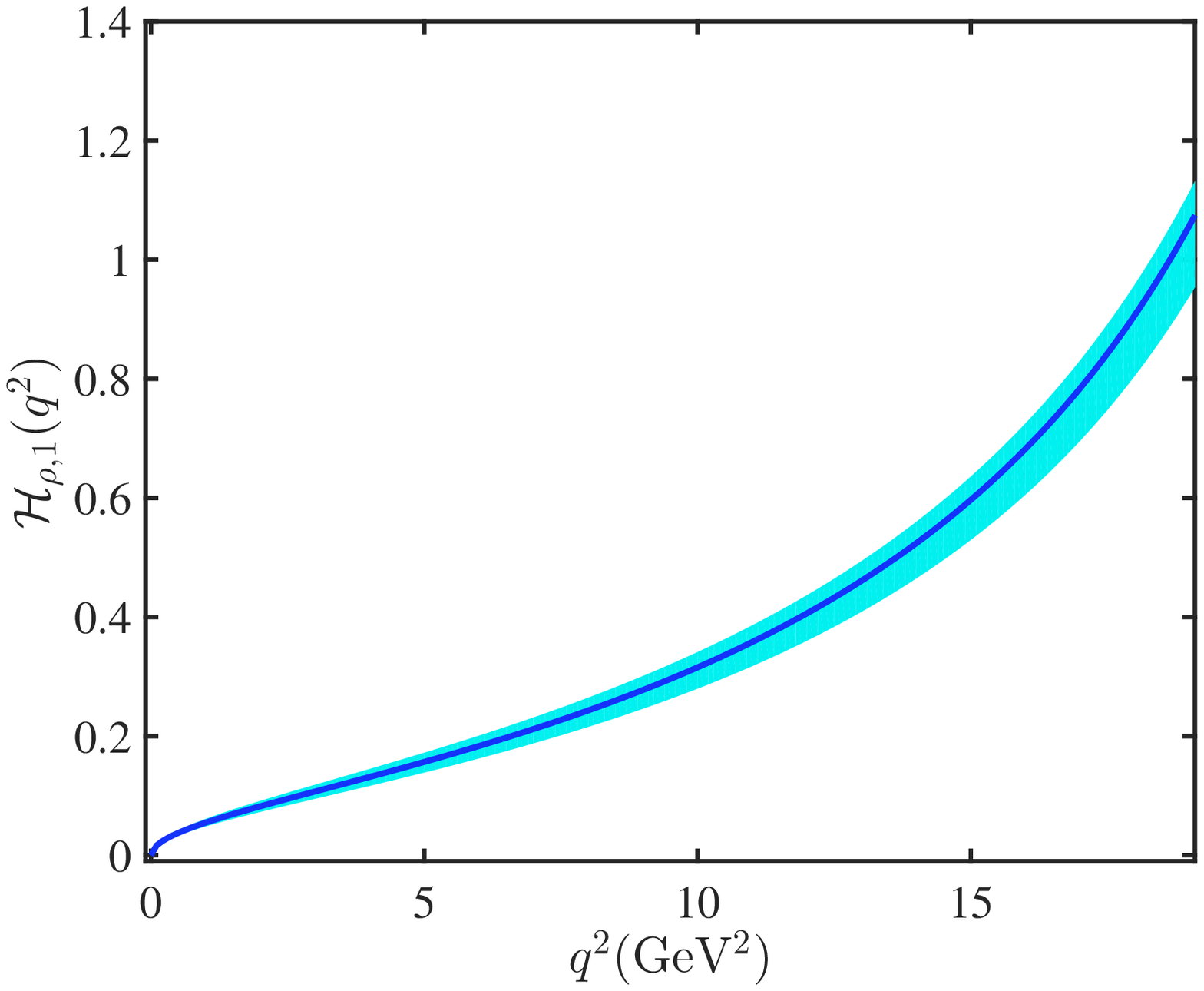}
\includegraphics[width=0.45 \textwidth, height=0.25 \textwidth]{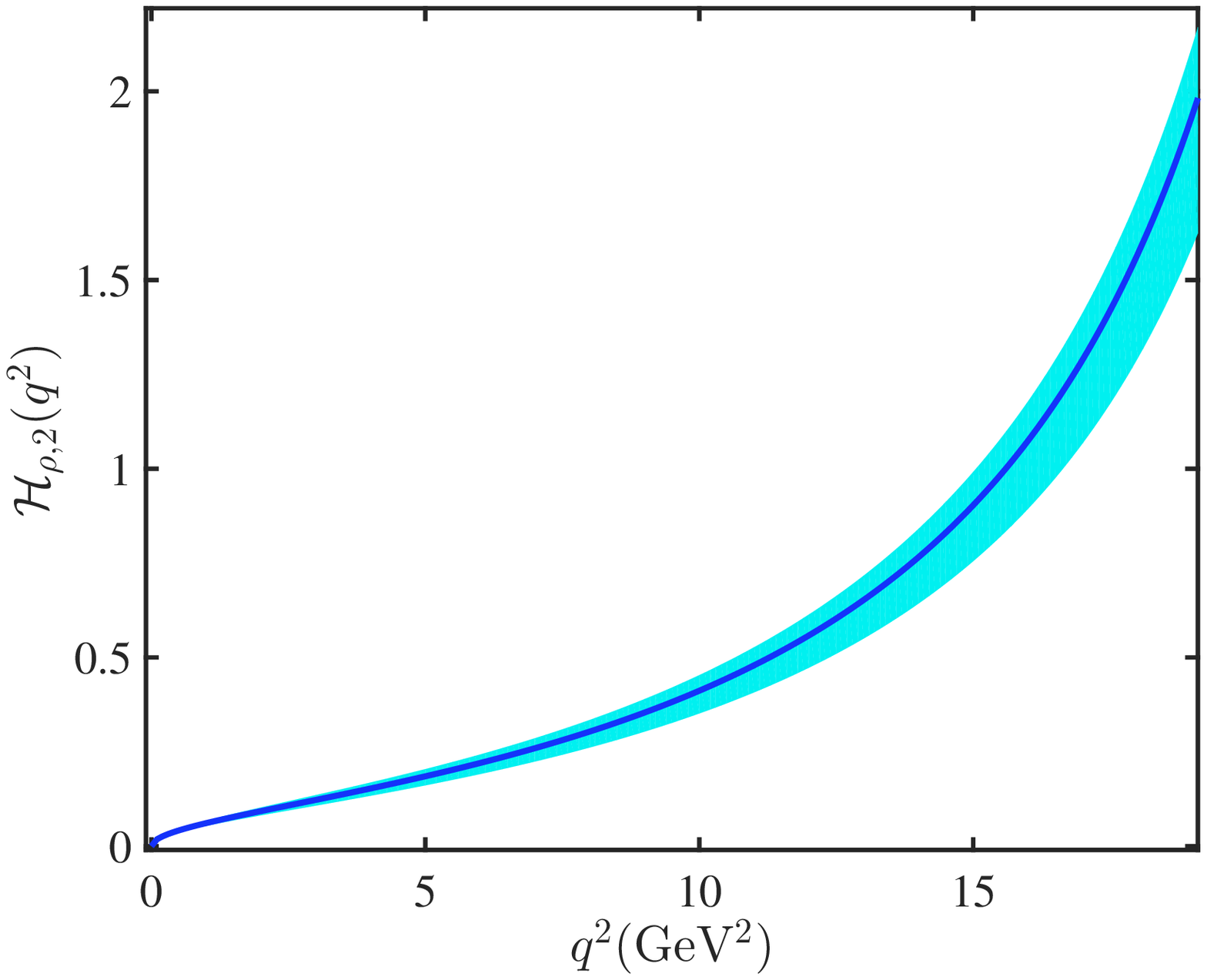}
\caption{The extrapolated LCSR predictions on the $B\to \rho$ HFFs $\mathcal{H}_{\rho,{(0,1,2)}}(q^2)$. The solid lines are center values and the shaded bands represent their uncertainties. }
\label{HFF:H012}
\end{center}
\end{figure}

We put the extrapolated $B\to \rho$ HFFs $\mathcal{H}_{\rho,\sigma}(q^2)$ in Fig.(\ref{HFF:H012}), where the shaded band stands for the squared average of all the mentioned uncertainties. All the HFFs are monotonically increase with the increment of $q^2$, and at the large recoil point, we have  $\mathcal{H}_{\rho,0}(0)=0.435^{+0.055}_{-0.045}$ and $\mathcal{H}_{\rho,\{1,2\}}(0)\equiv 0$.

\section{The $B\to \rho$ semileptonic decay and the CKM matrix element $|V_{\rm ub}|$}

In this subsection, we apply the HFFs $\mathcal{H}_{\rho,\sigma}(q^2)$ to study the semileptonic decay $B\to\rho\ell\nu_\ell$, which is frequently used for precision test the SM and for searching of new physics beyond SM.

\begin{figure}[htb]
\begin{center}
\includegraphics[width=0.45\textwidth]{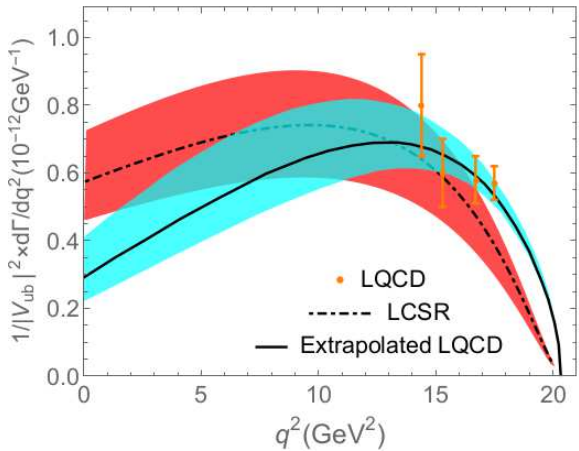}
\caption{The LCSR prediction for the differential decay width $1/|V_{ub}|^{2} \times d \Gamma/{d q^2}$. The LQCD prediction~\cite{Lattice96:1} and the extrapolated prediction of UKQCD group by using of the LQCD result~\cite{DelDebbio:1997nu} are presented as a comparison. The shaded bands are their theoretical errors. } \label{dGamma}
\end{center}
\end{figure}

Within the SM, the total differential decay width of $B\to\rho\ell\nu_\ell$ can be written as
\begin{equation}
\frac{1}{|V_{\rm ub}|^2}\frac{d\Gamma}{dq^2}={\cal G} \lambda(q^2)^{3/2} [\mathcal{H}_{\rho,0}^2(q^2) +\mathcal{H}_{\rho,1}^2(q^2)+\mathcal{H}_{\rho,2}^2(q^2)], \label{difftot}
\end{equation}
where the terms proportional $m_{\ell}^2$ have been suppressed due to the large chiral suppression for the light leptons with negligible masses, the parameter ${\cal G}={G_F^2}/{(192\pi^3 m_B^3)}$ with the fermi coupling constant $G_F=1.166\times10^{-5}~{\rm GeV}^{-2}$~\cite{Agashe:2014kda}, and the phase-space factor $\lambda(q^2)=(m_B^2 + m_\rho^2 - q^2)^2-4 m_B^2 m_\rho^2$. Our LCSR prediction for the differential decay width $1/|V_{\rm ub}|^{2} \times d \Gamma/{d q^2}$ is presented in Fig.(\ref{dGamma}), where the uncertainties from all error sources are added in quadrature. As a comparison, the UKQCD group LQCD prediction~\cite{Lattice96:1} and their extrapolated LQCD prediction (with the help of the heavy quark symmetry, kinematic constraints and the LCSR scaling relations)~\cite{DelDebbio:1997nu} are presented as a comparison. Our LCSR prediction is consistent with the LQCD prediction within the intermediate $q^2$-region; however our LCSR prediction prefer a larger $1/|V_{\rm ub}|^{2} \times d \Gamma/{d q^2}$ in low $q^2$-region and a smaller $1/|V_{\rm ub}|^{2} \times d \Gamma/{d q^2}$ in high $q^2$-region.

\begin{figure}[htb]
\begin{center}
\includegraphics[width=0.45\textwidth]{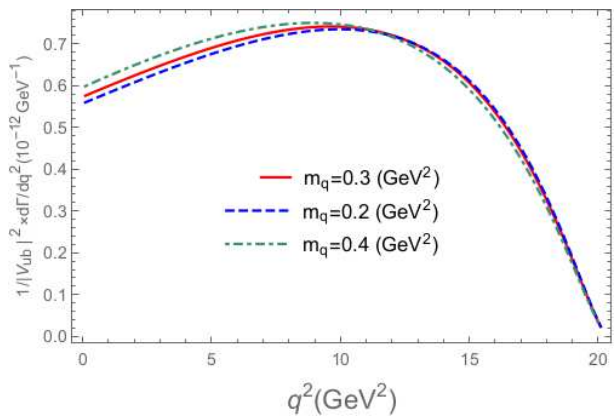}
\caption{The LCSR prediction for the differential decay width $1/|V_{ub}|^{2} \times d \Gamma/{d q^2}$ for $m_q\in[0.2, 0.4]$ GeV, where other input parameters are set to be their central values.} \label{dGammamq}
\end{center}
\end{figure}

As a minor point, we pick out the uncertainty caused by varying $m_q\in[0.2, 0.4]$ GeV from the above uncertainty, and present the LCSR prediction for the differential decay width $1/|V_{ub}|^{2} \times d \Gamma/{d q^2}$ in Fig.(\ref{dGammamq}). It shows the uncertainty caused by $m_q$ is small, which agree with the observation of Table~\ref{HFF uncertainties} that the dominant uncertainties are from the parameters $m_b$, $f_B$, and $s_0$.

\begin{table}[htb]
\caption{The LCSR predictions and the extrapolated LQCD predictions of the UKQCD group~\cite{DelDebbio:1997nu} for the total decay width $\Gamma/|V_{\rm ub}|^{2}$ and the ratio $\Gamma_{\|}/\Gamma_{\bot}$. }
\begin{tabular}{ c c c c c  }
\hline
&$\Gamma/|V_{ub}|^{2}$ & $\Gamma_{\|}/\Gamma_{\bot}$
\\ \hline
LCSR    & $12.1^{+2.6}_{-2.5}$ & $1.14^{+0.35}_{-0.34}$
\\
UKQCD   & $10.9^{+2.3}_{-1.5}$ & $0.80^{+0.04}_{-0.03}$
\\ \hline
\end{tabular}
\label{Ta:gammaTLS}
\end{table}

We present the total decay width $\Gamma/|V_{ub}|^{2}$ in Table \ref{Ta:gammaTLS}, in which we also present the ratio $\Gamma_{\|}/\Gamma_{\bot}$ as a useful reference. The total decay width, $\Gamma=\Gamma^\|+\Gamma^\bot$, where the decay width for the $\rho$-meson longitudinal components $\Gamma^\|$ is defined as
\begin{displaymath}
\Gamma^\|= {\cal G} |V_{\rm ub}|^2 \int_0^{q^2_{\rm max}} dq^2 \lambda(q^2)^{3/2} \mathcal{H}_{\rho,0}^2(q^2)
\end{displaymath}
and the decay width for the $\rho$-meson transverse components $\Gamma^\bot$ is defined as
\begin{displaymath}
\Gamma^\bot={\cal G} |V_{\rm ub}|^2 \int_0^{q^2_{\rm max}} dq^2 \lambda(q^2)^{3/2} [\mathcal{H}_{\rho,1}^2(q^2)+ \mathcal{H}_{\rho,2}^2(q^2)].
\end{displaymath}
Table \ref{Ta:gammaTLS} shows that, due to the large cancelation of the differences among different $q^2$-regions, the difference for the total decay width $\Gamma$ between the integrated LCSR and LQCD predictions shall be greatly suppressed.

\begin{figure}[htb]
\begin{center}
\includegraphics[width=0.45\textwidth]{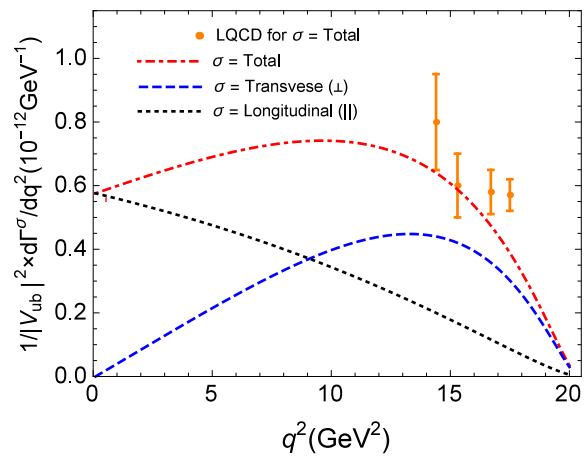}
\caption{The LCSR predictions for the polarized differential decay widths $1/|V_{ub}|^{2} \times d \Gamma^{\|}/{d q^2}$ and $1/|V_{ub}|^{2} \times d \Gamma^{\bot}/{d q^2}$. The LQCD result for total differential decay width~\cite{Lattice96:1} is presented as a comparison. } \label{dGammapol}
\end{center}
\end{figure}

We present the LCSR predictions for the polarized differential decay widths $1/|V_{ub}|^{2} \times d \Gamma^{\|}/{d q^2}$ and $1/|V_{ub}|^{2} \times d \Gamma^{\bot}/{d q^2}$ in Fig.(\ref{dGammapol}), in which all the input parameters are set to be their central values. Fig.(\ref{dGammapol}) shows that the differential decay widths for the final-state $\rho$-meson transverse and longitudinal components behave quite differently. The longitudinal differential decay width $d \Gamma^{\|}/{d q^2}$ monotonously deceases with the increment of $q^2$, and the transverse differential decay width $d \Gamma^{\bot}/{d q^2}$ shall first increase and then decrease with the increment of $q^2$. Both of them tend to zero for $q^2\to q^2_{\rm max}$ due to the phase-space suppression. As a result, the $\rho$-meson longitudinal component dominates low $q^2$-region, and its transverse component dominates high $q^2$-region~\footnote{Such dominance could be explained as a consequence of Lorentz invariance~\cite{Hiller:2013cza}.}.

\begin{figure}[htb]
\begin{center}
\includegraphics[width=0.45\textwidth]{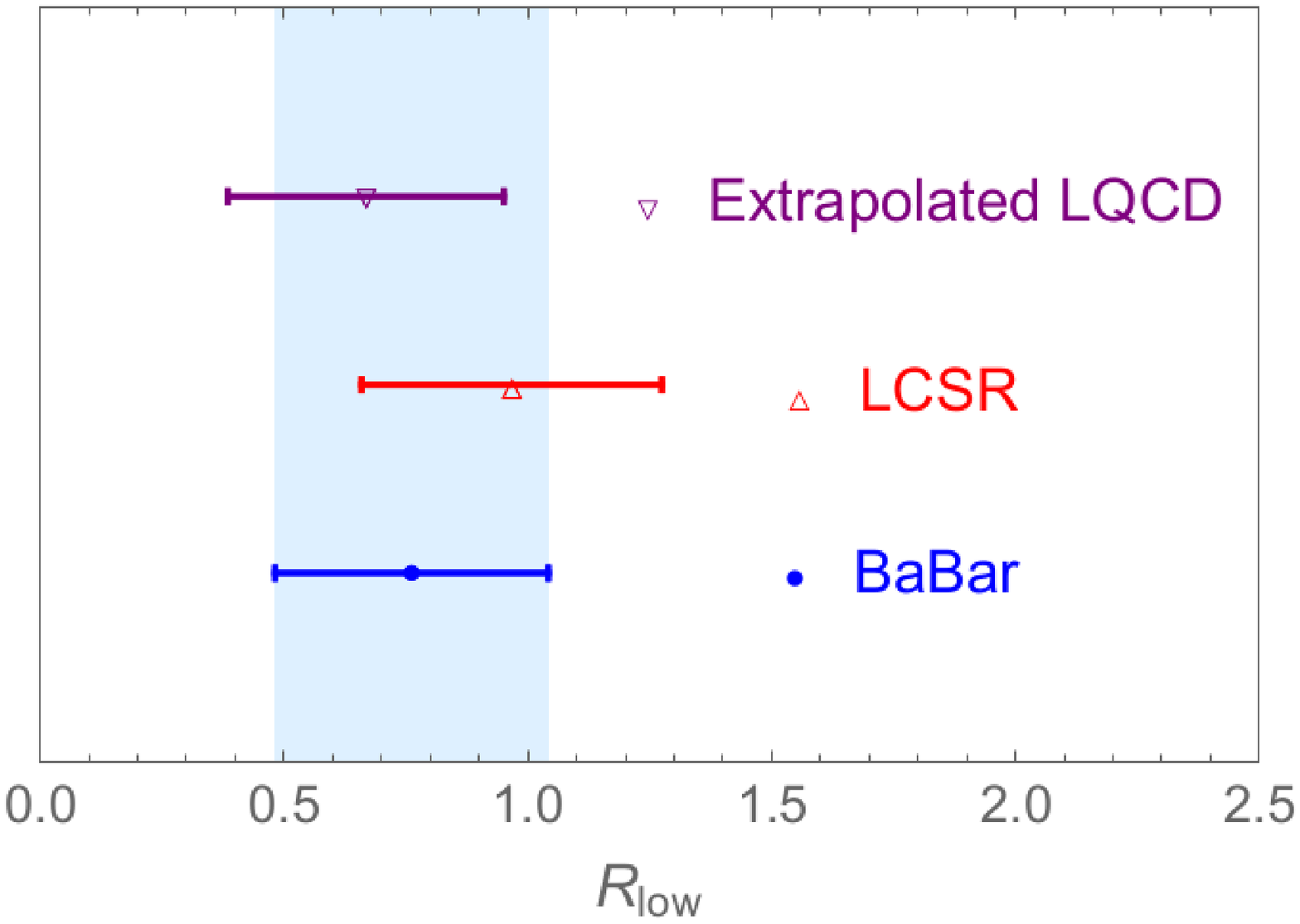}
\includegraphics[width=0.45\textwidth]{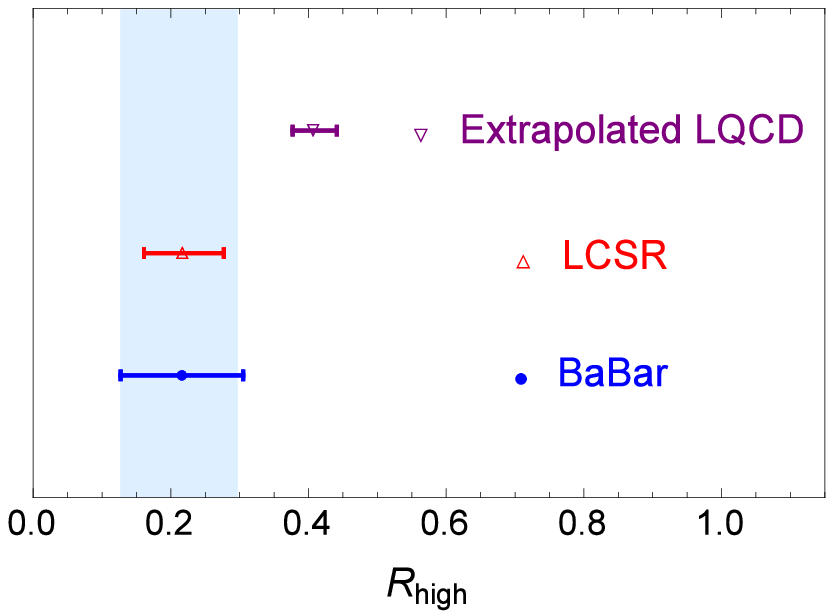}
\caption{The LCSR predictions for the ratios $R_{\rm low}$ and $R_{\rm high}$. The BaBar~\cite{delAmoSanchez:2010af} results and the values by using extrapolated LQCD predictions~\cite{DelDebbio:1997nu} are also presented. }
\label{Rlowhigh}
\end{center}
\end{figure}

Experimentally, the BaBar collaboration measured the partial decay widths in three different $q^2$-regions~\cite{delAmoSanchez:2010af}
\begin{eqnarray}
\Delta \Gamma_{\rm low} &=& \int_0^8 \frac{d \Gamma}{d q^2} d q^2 = (0.747 \pm 0.234) \times 10^{-4}, \\
\Delta \Gamma_{\rm mid} &=& \int_8^{16} \frac{d \Gamma}{d q^2} d q^2 = (0.980 \pm 0.187) \times 10^{-4}, \\
\Delta \Gamma_{\rm high}&=& \int_{16}^{20.3} \frac{d \Gamma}{d q^2} d q^2 = (0.256 \pm 0.072) \times 10^{-4},
\end{eqnarray}
which lead to
\begin{eqnarray}
R_{\rm low}  &=& \frac{\Gamma_{\rm low}}{\Gamma_{\rm mid}}  = 0.762 \pm 0.280, \label{EQlow}  \\
R_{\rm high} &=& \frac{\Gamma_{\rm high}}{\Gamma_{\rm mid}} = 0.216 \pm 0.089. \label{EQhigh}
\end{eqnarray}
Our LCSR calculation gives, $R_{\rm low}=0.967^{+0.308}_{-0.285}$ and $R_{\rm high}=0.219^{+0.058}_{-0.070}$; and the extrapolated LQCD calculation gives, $R_{\rm low}=0.668^{+0.283}_{-0.154}$ and $R_{\rm high}=0.409^{+0.032}_{-0.051}$. A comparison of those two ratios is presented in Fig.(\ref{Rlowhigh}). The LCSR predictions agree with the BaBar measurement with errors, while the extrapolated LQCD prefers a larger $R_{\rm high}$, which is about $1.6\,\sigma$ deviation from the BaBar measurement. Because the (middle) partial decay widths $\Delta \Gamma_{\rm mid}$ for the LCSR and LQCD  approaches are close to each other, by comparing $R_{\rm low}$ and $R_{\rm high}$ with the experimental data, one can get the correct decay widths in different $q^2$-region and thus confirm which theoretical prediction is more reliable.

As a final remark, with the help of the branching ratio ${\cal B}(B^0\to\rho^-\ell^+\nu_\ell)=(2.45\pm 0.32)\times 10^{-4}$ and the lifetime $\tau(B^0)= 1.520\pm 0.004 {\rm ps}$~\cite{Beringer:1900}, we obtain $|V_{\rm ub}|=(2.96^{+0.52}_{-0.51})\times10^{-3}$, where the error is weighted average of all the mentioned error sources. This value agrees with the BaBar predictions~\cite{Wulsin:2010fp}, $(2.75\pm 0.24)\times10^{-3}$ and $(2.83\pm 0.24)\times10^{-3}$, and the CLEO predictions~\cite{Behrens:1999vv}, $(3.23\pm 0.24^{+0.23}_{-0.26}\pm0.58)\times10^{-3}$ and $(3.25\pm 0.14^{+0.21}_{-0.29}\pm0.55)\times10^{-3}$, within errors.

\section{Summary} \label{summary}

We have studied the HFFs for the $B$-meson semileptonic decay $B\to\rho\ell\nu_\ell$ within the LCSR approach. Fig.(\ref{HFF:H012}) shows that the extrapolated HFFs within the whole $q^2$-region. At the large recoil point, only the $\rho$-meson longitudinal component contributes, e.g. $\mathcal{H}_{\rho,0}(0)=0.435^{+0.055}_{-0.045}$ and $\mathcal{H}_{\rho,\{1,2\}}(0)\equiv 0$, where the errors are squared averages of the considered error sources. By applying the extrapolated HFFs to the semileptonic decay $B\to\rho\ell\nu_\ell$, we observe that the differential decay width $1/|V_{\rm ub}|^{2} \times d \Gamma/{d q^2}$, as shown by Fig.(\ref{dGamma}), is consistent with the Lattice QCD prediction within the intermediate $q^2$-region. However our LCSR prediction prefer a larger $1/|V_{\rm ub}|^{2} \times d \Gamma/{d q^2}$ in low $q^2$-region and a smaller $1/|V_{\rm ub}|^{2} \times d \Gamma/{d q^2}$ in high $q^2$-region. More explicitly, Fig.(\ref{dGammapol}) shows that the longitudinal decay width dominates the lower $q^2$-region and the transverse one dominates the higher $q^2$-region. Two typical ratios $R_{\rm low}$ and $R_{\rm high}$ can be used to test those properties. Our LCSR calculation shows that $R_{\rm low}=0.967^{+0.308}_{-0.285}$ and $R_{\rm high}=0.219^{+0.058}_{-0.070}$. Fig.(\ref{Rlowhigh}) shows that those predictions agree with the BaBar measurements within errors. Thus by using the HFFs with definite polarizations, some useful information can be achieved. A more precise measurement of those ratios shall be helpful for testing various calculation approaches.

\hspace{2cm}

{\bf Acknowledgements}: This work was supported in part by the Natural Science Foundation of China under Grant No.11625520 and No.11765007; by the Fundamental Research Funds for the Central Universities under the Grant No.2018CDPTCG0001/3; by the Project of Guizhou Provincial Department of Science and Technology under Grant No.[2017]1089; by the Project for Young Talents Growth of Guizhou Provincial Department of Education under Grant No.KY[2016]156; the Key Project for Innovation Research Groups of Guizhou Provincial Department of Education under Grant No.KY[2016]028.

\appendix

\section*{Appendix: the nonlocal matrix elements}

The  nonlocalmatrix elements used in our calculation are~\cite{Ball:2004rg, Ball:2007zt, Fu:2014uea}:
\begin{widetext}
\begin{eqnarray}
\langle \rho(k,\varepsilon(k))|\bar d(x) q(0)|0\rangle &=& - \frac{i}{2} f_\rho^\bot(E\cdot x) m_\rho^2\int_0^1 du e^{iup\cdot x} \psi_{3;\rho }^\parallel(u),\\
\langle \rho(k,\varepsilon(k))|\bar d(x) \gamma _\beta \gamma _5 q(0)|0\rangle  &=& \frac{1}{4}\varepsilon _\beta m_\rho f_\rho^ \parallel \int_0^1 d u e^{iup\cdot x}\psi_{3;\rho}^\bot(x),\\
\langle \rho(k,\varepsilon(k))|\bar d(x)\gamma _\beta q(0)|0\rangle  &=& m_\rho f_ \rho ^\parallel \int_0^1 d ue^{iu  p\cdot x} \bigg\{ \frac{E\cdot x}{p \cdot x} p_\beta  \phi _{2;\rho}^ \parallel (u)  +  E_\beta \phi _{3;\rho}^ \bot (u)\bigg\},\\
\langle \rho(k,\varepsilon(k))|\overline d(x) \gamma _\beta q (0)|0\rangle  &=& m_\rho f_\rho^\parallel \int_0^1 d u e^{iu p\cdot x} \bigg\{ \frac{E \cdot x}{{p \cdot x}}{p_\beta }\left[ {\phi _{2;\rho}^\parallel (u) + \phi _{3;{\rho}}^ \bot (u)} \right] + \frac{E \cdot x}{{p \cdot x}}{p_\beta }\frac{m_\rho ^2 x^2}{16}\phi _{4;\rho}^\parallel (u)\nonumber\\
&&+ E_\beta \phi _{3;\rho}^ \bot (u) - \frac{1}{2} x_\beta \frac{E \cdot x}{(p \cdot x)^2}m_\rho^2 \bigg[ \psi _{4;\rho}^\parallel (u) + \phi _{2;\rho}^\parallel (u) - 2 \phi _{3;\rho}^\bot (u) \bigg]\bigg\},\\
\langle \rho (p,\lambda)|\bar d(x)\sigma_{\mu \nu}q(0)|0\rangle &=& - i f_\rho^\bot \int_0^1 du e^{iu p\cdot x}\bigg\{(E_\mu p_\nu - E_\nu p_\mu)\bigg[\phi_{2;\rho}^\bot(u)+ \frac{m_\rho^2 x^2}{16} \phi_{4;\rho}^\bot(u)\bigg]  \nonumber\\
&& +\left(p_\mu x_\nu - p_\nu x_\mu \right) \frac{E \cdot x}{(p\cdot x)^2} m_\rho^2 \bigg[\phi_{3;\rho}^\|(u) - \frac{1}{2} \phi_{2;\rho}^\bot(u) - \frac{1}{2} \psi_{4;\rho}^\bot(u)\bigg] \nonumber\\
&& +\frac{1}{2} \left( E_\mu x_\nu  - E_\nu x_\mu \right) \frac{m_\rho^2}{p\cdot x}\bigg[ \psi_{4;\rho}^\bot(u)-\phi_{2;\rho}^\bot(u) \bigg] \bigg\},\label{DA1}
\end{eqnarray}
\end{widetext}
where $f_\rho^\bot$ and $f_\rho^\|$ are $\rho$-meson decay constants, which are defined as
\begin{eqnarray}
\left\langle \rho(k,\varepsilon(k))\right|\bar d(0){\gamma _\mu }q(0)\left| 0 \right\rangle  &=& f_{\rho}^\parallel {m_{{\rho}}}E_\mu,\\
\left\langle \rho(k,\varepsilon(k))\right|\bar d(0){\sigma _{\mu \nu }}q(0)\left| 0 \right\rangle &=& if_{\rho}^ \bot (E_\mu {p_\nu } - E_\nu {p_\mu }).
\end{eqnarray}
To do the simplification, the following identities are helpful:
\begin{eqnarray}
\gamma _\mu \gamma _\nu  &=& g_{\mu \nu } - i\sigma _{\mu \nu }, \\
\gamma _\mu \gamma _\nu \gamma _5 &=& g_{\mu \nu }\gamma _5 - \frac{1}{2}\varepsilon _{\mu \nu \alpha \beta }\sigma ^{\alpha \beta },\\
\gamma _5 \sigma ^{\rho \sigma } &=&  - \frac{i}{2}\sigma ^{\alpha \beta }\varepsilon _{\rho \sigma \alpha \beta },\\
\sigma _{\mu \nu }\gamma ^\alpha &=& i(g_{\alpha \nu }\gamma _\mu  - g_{\alpha \mu }\gamma _\nu ) + \varepsilon _{\alpha \mu \nu \beta }\gamma ^\beta \gamma ^5.
\end{eqnarray}

\end{document}